\documentclass[sigconf, screen]{acmart}

\AtBeginDocument{%
  }

\setcopyright{acmlicensed}
\copyrightyear{2027}
\acmYear{2027}
\acmDOI{XXXXXXX.XXXXXXX}
\acmConference[WWW Companion '27]{
ACM Web Conference 2027}{May 10--14, 2027}{Dublin, Ireland}
\acmBooktitle{
ACM Web Conference 2027, 
May 10--14, 2027, Dublin, Ireland}
\acmISBN{978-1-4503-XXXX-X/2027/05}

\usepackage{xspace}
\usepackage{enumitem}
\usepackage{tikz}
\usepackage{pgfplots}
\usepackage{acronym}

\acrodef{EEG}{electroencephalogram}
\acrodef{BCI}{brain-computer interface}
\acrodef{AI}{Artificial intelligence}
\acrodef{FAA}{Frontal Alpha Asymmetry}
\acrodef{AI}{Arousal Index}
\acrodef{LOF}{Local Outlier Factor}

\usetikzlibrary{calc,positioning}
\pgfplotsset{compat=1.18}

\newcommand{\va}{V/A\xspace}

\newcommand{\citemissing}[1][]{\textcolor{red}{[CITE\ifx&#1&\else: #1\fi]}}

\begin{document}

\title{EE-Motion: An EEG-based Web Architecture for Live Affective Music Recommendation}

\author{Sabino Roccotelli}
\affiliation{%
  \institution{Politecnico di Bari}
  \city{Bari}
  \country{Italy}}
\authornote{Corresponding authors. These authors contributed equally to this research.}

\author{Potito Aghilar}
\authornotemark[1]
\affiliation{%
  \institution{Politecnico di Bari}
  \city{Bari}
  \country{Italy}}

\author{Vito Walter Anelli}
\authornotemark[1]
\affiliation{%
  \institution{Politecnico di Bari}
  \city{Bari}
  \country{Italy}}

\author{Alejandro Bellog\'{\i}n}
\affiliation{%
  \institution{Univ. Aut\'{o}noma de Madrid}
  \city{Madrid}
  \country{Spain}}

\author{Tommaso Colafiglio}
\affiliation{%
  \institution{Politecnico di Bari}
  \city{Bari}
  \country{Italy}}

\author{Michelantonio Trizio}
\affiliation{%
  \institution{Wideverse}
  \city{Bari}
  \country{Italy}}

\author{Tommaso Di Noia}
\affiliation{%
  \institution{Politecnico di Bari}
  \city{Bari}
  \country{Italy}}

\renewcommand{\shortauthors}{Roccotelli et al.}
\settopmatter{authorsperrow=4}

\begin{abstract}
Music listening is intrinsically tied to emotion, which drifts continuously within a single session. Although modern browsers interface with wearable brain-computer interfaces, recommendation systems remain structured around discrete logs and lack the infrastructure to integrate continuous physiological signals. Furthermore, existing affect-aware models rely on static user profiles and offline training, failing to serve anonymous newcomers without model retraining. We present a web platform that resolves both challenges. Pairing with a consumer \ac{EEG} headband over Web Bluetooth, the system captures live \ac{EEG}, mapping frontal alpha asymmetry and arousal index to normalized valence-arousal proxy coordinates on Russell's circumplex plane. Driven by these coordinates, an affective adapter unifies structurally diverse recommendation backbones under a shared live scoring mechanism. By parameterizing anonymous visitors through a few favorite tracks, the module recommends tracks conditioned on live proxy coordinates. This demonstrates that real-time EEG-derived control signals can steer recommendations without persistent user tables or model retraining. A live demo and video walkthrough are available \href{https://ee-motion.github.io/}{\textit{here}}.


\end{abstract}




\maketitle

\acresetall

\section{Introduction}
\label{sec:introduction}

\begin{figure*}[t]
  \centering
  \includegraphics[width=\linewidth]{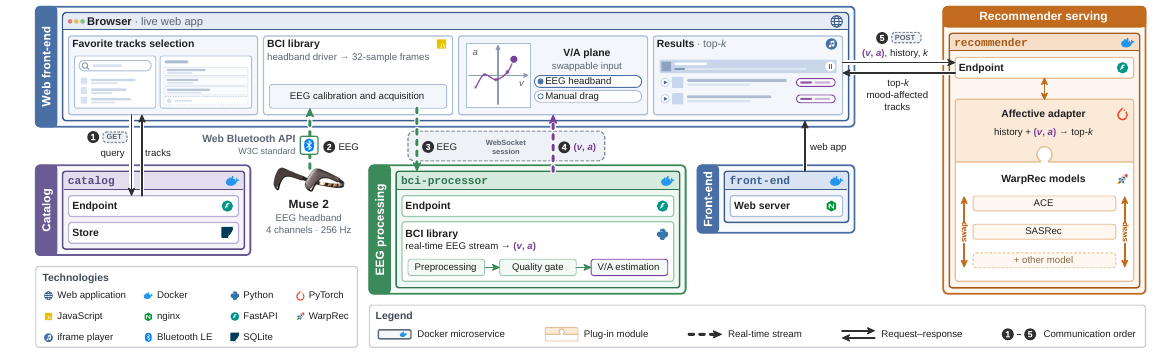}
  \caption{Overview of the Web-based Live Affective Architecture. The browser acquires raw EEG via Web Bluetooth and streams it to the \textit{bci-processor} for \va\ mapping; the \textit{affective adapter} then scores candidates across swappable WarpRec backbones.}
  \label{fig:architecture}
\end{figure*}

Music listening experiences are intrinsically governed by a listener's affective state, which is never static but continuously drifts within a single session~\cite{DBLP:journals/corr/abs-2508-20289}. In affective computing and \ac{BCI} research, this state is commonly described by the continuous valence and arousal (\va) dimensions of Russell's circumplex model~\cite{DBLP:journals/taffco/AlarcaoF19} (measuring emotional positivity and physiological activation, respectively). Indeed, emotion-elicitation studies show that listening to even a single minute of music reliably induces distinct \va shifts~\cite{DBLP:journals/taffco/AlarcaoF19}, meaning a listener's mood can fluctuate multiple times throughout a listening session.
Despite this continuous drift, contemporary recommender systems remain largely blind to real-time emotional shifts. When systems incorporate affect, they typically infer it from past listening logs, static context attributes, or explicit self-reports~\cite{DBLP:journals/corr/abs-2508-20289}. Recent models such as ACE~\cite{DBLP:conf/sigir/AghilarRABTN26}, for instance, anchor each user to the \va centroid of previously consumed items. While effective in offline benchmarks, such static anchors remain fixed even when a listener's mood departs from their historical average during an 
active session. 

Aligning recommendations with a listener's immediate affective state requires capturing this signal during the session and passing it to the ranking pipeline at request time. In practice, establishing this live closed loop exposes three deeply connected operational and methodological bottlenecks across the system pipeline.

(i) First, at the serving infrastructure level, production recommendation systems are engineered around episodic logs of discrete actions, such as clicks, plays, and ratings~\cite{DBLP:reference/sp/KorenRB22}. While context-aware models incorporate situational factors~\cite{DBLP:reference/sp/AdomaviciusT15}, they typically assume discrete attributes known ahead of time. In contrast, continuous physiological telemetry sampled at hundreds of hertz in the browser is noisy and susceptible to artifacts, requiring real-time filtering and feature extraction before yielding a stable affective estimate~\cite{DBLP:journals/tcbb/GuCJXWJL21}. Even modern serving frameworks designed for low-latency ranking, such as WarpRec~\cite{DBLP:journals/corr/abs-2602-17442}, provide no native hook to forward continuous in-browser telemetry directly to custom ranking models.

(ii) Second, at the recommendation model level, streaming this affective signal addresses only half the challenge, because the visitor in front of the screen is typically an anonymous newcomer. Collaborative filtering models allocate static user embedding tables that require expensive retraining or per-user fold-in to represent new visitors, unless users are parameterized purely through consumed items~\cite{DBLP:reference/sp/KorenRB22}. Sequential architectures, such as self-attentive transformers (e.g., SASRec~\cite{DBLP:conf/icdm/KangM18}), elegantly bypass this user-table bottleneck by encoding an ordered sequence of interactions directly into a dynamic user state without persistent user identifiers. Yet, because sequential models operate strictly over discrete item tokens, they lack native mechanisms to fuse external, continuous affective factors during candidate scoring. Compounding this divide, existing affect-aware techniques bind continuous \va signals directly to model-specific training objectives~\cite{DBLP:journals/corr/abs-2508-20289,DBLP:conf/sigir/AghilarRABTN26}, leaving researchers without a shared abstraction to project live affect into the recommendation process across structurally different models.



(iii) Third, \ac{BCI}s have historically remained isolated from the web ecosystem. While implicit affect sensing has explored video feeds to detect emotion~\cite{DBLP:journals/corr/abs-2508-20289,DBLP:journals/tmm/TkalcicOKT13}, wearable \ac{EEG} headsets are able to record raw brain signals from which affective states can be inferred ~\cite{DBLP:journals/taffco/AlarcaoF19,DBLP:journals/tcbb/GuCJXWJL21}.
Although web-based streaming is increasingly explored, accessing wearable \ac{BCI} telemetry has traditionally relied on third-party drivers. The emerging W3C Web Bluetooth specification\footnote{\url{https://webbluetoothcg.github.io/web-bluetooth/}} removes this barrier, enabling browsers to pair directly with Bluetooth Low Energy hardware without local installation.
Yet, the web ecosystem still lacks an open, end-to-end architecture to translate continuous in-browser bio-signals into live recommendation context.


To resolve these challenges within a unified system, we present a web platform that connects in-browser bio-sensing directly to real-time music recommendation. The platform integrates two decoupled core components: (1) an in-browser acquisition pipeline that pairs with a consumer \ac{EEG} device over Web Bluetooth, preprocesses the raw bio-signals in real time, and maps them to normalized valence and arousal proxy coordinates; and (2) an affective adapter that unifies candidate scoring across different model families. This adapter allows anonymous visitors to interact with the system by providing a small set of favorite tracks that capture their musical preferences, while their live \ac{EEG} signal steers recommendations in real time without model retraining. Consequently, structurally different models, such as graph neural networks~\cite{DBLP:conf/sigir/AghilarRABTN26} and sequential recommendation~\cite{DBLP:conf/icdm/KangM18}, can be swapped at serving time without modifying the web client or serving infrastructure.

\section{System Architecture}
\label{sec:architecture}

\subsection{Overview}
\label{sec:overview}

Live affective music recommendation requires coordinating three distinct operational tasks: in-browser sensor communication, continuous V/A proxy estimation, and low-latency candidate scoring. The system addresses these demands through a decoupled microservice architecture that isolates candidate ranking from sensor-specific processing (Figure~\ref{fig:architecture}). Four independent services carry these responsibilities: (i) the \textit{front-end} serves the interactive web application to the client browser; (ii) the \textit{catalog} handles music track metadata retrieval; (iii) the \textit{bci-processor} is dedicated to real-time \ac{EEG} preprocessing and affective feature extraction; and (iv) the \textit{recommender} scores candidates across the catalog using WarpRec~\cite{DBLP:journals/corr/abs-2602-17442}.

Two core contributions bridge these independent services into an end-to-end pipeline. First, a standalone \textit{\ac{BCI} library} connects to the headband via Web Bluetooth and streams EEG data to the \textit{bci-processor}, which computes normalized \va proxy coordinates. Second, an \textit{affective adapter} hooks into WarpRec models with minimal implementation effort. In addition to live \va projection, this adapter transforms offline models of different families (such as graph neural networks or sequential architectures) into their live affective variants, serving anonymous visitors on the fly without retraining. This architectural separation keeps the recommendation model completely agnostic to the physical origin of the affective signal. 
The \textit{recommender} requires only normalized valence and arousal proxy coordinates in $[0,1]\times[0,1]$, obtained from \ac{FAA} and \ac{AI}~\cite{DBLP:journals/taffco/AlarcaoF19} via a fixed-coefficient sigmoid.
As long as an input source produces values within this normalized range, the ranking models function identically, regardless of the underlying sensing modality. This property directly unlocks what we call
\emph{input swappability}: the system seamlessly accepts any alternative input that emits the same coordinate format.
HTTP and WebSocket protocols decouple these microservices, providing deployment flexibility across heterogeneous topologies.
This modularity allows the platform to run either as a self-contained local edge deployment (as in our standalone demonstration) or as a distributed deployment across dedicated nodes. 

\subsection{Brain-Computer Interface Protocol}
\label{sec:bci_signal_processing}



The \ac{BCI} workflow operates across two distinct phases: acquisition and processing.
\ac{EEG} signals are acquired from a Muse~2 device\footnote{\url{https://choosemuse.com/products/muse-2}} using Muse-js,\footnote{\url{https://github.com/urish/muse-js}} with four channels (TP9, AF7, AF8, TP10) sampled at 256~Hz. In the \textit{bci-processor}, a dedicated thread maintains a circular buffer containing 1024 samples per channel, corresponding to 4~s of EEG.
Once the buffer is full, the acquired data are retrieved and converted into an MNE Raw data structure~\cite{gramfort2013mne}, where signal quality is assessed using MNE's \ac{LOF} channel detection function\footnote{\url{https://mne.tools/stable/generated/mne.preprocessing.find_bad_channels_lof.html}} to identify channels with anomalous signal characteristics. Visual inspection complements automated assessment, supporting noise identification and headset repositioning. Preprocessing then applies causal bandpass filtering between 1 and 40~Hz, after which NeuroKit2\footnote{\url{https://neuropsychology.github.io/NeuroKit/index.html}} estimates theta (4-8~Hz), alpha (8-13~Hz), and beta (13-30~Hz) powers. To attenuate short-term fluctuations, a second circular buffer stores 20 successive observations, allowing temporal smoothing to average each feature across these observations, separately for each band and channel. Affective indices are computed once this buffer is full: \ac{FAA} serves as the valence-related index, consistent with prior evidence, and is computed as $\log(\alpha_{\mathrm{AF8}}/\alpha_{\mathrm{AF7}})$, while the \ac{AI} is defined as $(\beta_{\mathrm{AF7}}+\beta_{\mathrm{AF8}})/(\alpha_{\mathrm{AF7}}+\alpha_{\mathrm{AF8}})$~\cite{shao2020affect}.
Here, $\alpha$ and $\beta$ denote temporally smoothed powers in the alpha and beta bands. \ac{FAA} and \ac{AI} serve as physiological proxy indices rather than validated emotion recognition.

\subsection{Affective Adapter}
\label{sec:adapter}

The \textit{affective adapter} operates inside the \textit{recommender} service as a decoupled module that bridges offline-trained ranking models with live affective serving (Figure~\ref{fig:architecture}). Its objective is to equip the recommendation backbones with a live-affect scoring mechanism, allowing them to serve anonymous web visitors without persistent user embeddings or online retraining. 
The design stems from the affective grounding principles introduced in ACE~\cite{DBLP:conf/sigir/AghilarRABTN26}. Specifically, we abstracted how ACE infuses valence and arousal into representation learning, generalizing its mechanics to support structurally different recommendation families. At serving time, the adapter operates on three inputs: (i) the tracks selected by the user, (ii) the live \va coordinates, and (iii) the pre-computed item embeddings, each associated with an item-level \va profile. In response, it outputs the top-$k$ ranked tracks, conditioned on the input V/A proxy coordinates.
This architecture resolves two fundamental bottlenecks of live deployment: anonymous visitor representation and model heterogeneity. First, traditional collaborative filtering models rely on static user embeddings that fail to handle unobserved web visitors~\cite {DBLP:reference/sp/KorenRB22}. 
The adapter overcomes this barrier by projecting an anonymous visitor into the latent space via averaging the embeddings of the visitor's selected tracks, bypassing user tables and eliminating online retraining~\cite{DBLP:reference/sp/KorenRB22,DBLP:conf/nips/ZaheerKRPSS17}.
Second, recommenders rely on fundamentally different scoring mechanics (e.g., SASRec~\cite{DBLP:conf/icdm/KangM18} and ACE~\cite{DBLP:conf/sigir/AghilarRABTN26}).
The adapter bridges this divide and makes ranking models fully interchangeable by exposing a serving interface that is based solely on the visitor's history and live \va coordinates. 

\subsection{Quantitative Profiling}
\label{sec:profiling}

\pgfplotstableread[col sep=comma]{data/trace.csv}\perfTrace
\pgfplotstableread[col sep=comma]{data/trace_cycle.csv}\perfCycle
\begin{figure}[t]
\centering
\begin{tikzpicture}[font=\sffamily,
    tracevalue/.style={anchor=west, inner sep=0pt, font=\fontsize{6}{7}\selectfont, text=gray!80!black},
    cyclevalue/.style={anchor=west, inner sep=0pt, font=\fontsize{6}{7}\selectfont\bfseries}]

\definecolor{cDsp}{RGB}{30, 123, 110}    
\definecolor{cRec}{RGB}{200, 90, 23}     
\definecolor{cCat}{RGB}{112, 76, 160}    
\definecolor{cCyc}{RGB}{110, 110, 110}   
\definecolor{cGrid}{RGB}{242, 242, 242}  

\providecommand{\tracerow}[2]{%
    \pgfplotstablegetelem{#1}{y}\of\perfTrace\let\tY\pgfplotsretval
    \pgfplotstablegetelem{#1}{start}\of\perfTrace\let\tS\pgfplotsretval
    \pgfplotstablegetelem{#1}{e1}\of\perfTrace\let\tA\pgfplotsretval
    \pgfplotstablegetelem{#1}{e2}\of\perfTrace\let\tB\pgfplotsretval
    \pgfplotstablegetelem{#1}{e3}\of\perfTrace\let\tE\pgfplotsretval
    \pgfplotstablegetelem{#1}{rtt_sd}\of\perfTrace\let\tSd\pgfplotsretval
    \pgfplotstablegetelem{#1}{send}\of\perfTrace\let\tSend\pgfplotsretval
    \pgfplotstablegetelem{#1}{proc}\of\perfTrace\let\tProc\pgfplotsretval
    \pgfplotstablegetelem{#1}{recv}\of\perfTrace\let\tRecv\pgfplotsretval
    \pgfmathsetmacro\tLo{\tY-0.24}\pgfmathsetmacro\tHi{\tY+0.24}%
    \pgfmathsetmacro\tWlo{max(\tS, \tE-\tSd)}\pgfmathsetmacro\tWhi{\tE+\tSd}%
    \edef\temp{%
        \noexpand\fill[#2!30] (axis cs:\tS,\tLo) rectangle (axis cs:\tA,\tHi);
        \noexpand\fill[#2] (axis cs:\tA,\tLo) rectangle (axis cs:\tB,\tHi);
        \noexpand\fill[#2!30] (axis cs:\tB,\tLo) rectangle (axis cs:\tE,\tHi);
        \noexpand\draw[#2, line width=0.4pt] (axis cs:\tS,\tLo) rectangle (axis cs:\tE,\tHi);
        \noexpand\whisker{\tWlo}{\tWhi}{\tY}
        \noexpand\node[tracevalue] at ({rel axis cs:1.03,0} |- {axis cs:0,\tY})
            {(\noexpand\pgfmathprintnumber[fixed, fixed zerofill, precision=1, assume math mode=true]{\tSend} + \noexpand\pgfmathprintnumber[fixed, fixed zerofill, precision=1, assume math mode=true]{\tProc}
             + \noexpand\pgfmathprintnumber[fixed, fixed zerofill, precision=1, assume math mode=true]{\tRecv}) \noexpand\textpm{} \noexpand\pgfmathprintnumber[fixed, fixed zerofill, precision=1, assume math mode=true]{\tSd}};}%
    \temp}
\providecommand{\tracecycle}[1]{%
    \pgfplotstablegetelem{#1}{y}\of\perfCycle\let\tY\pgfplotsretval
    \pgfplotstablegetelem{#1}{rtt}\of\perfCycle\let\tE\pgfplotsretval
    \pgfplotstablegetelem{#1}{rtt_sd}\of\perfCycle\let\tSd\pgfplotsretval
    \pgfmathsetmacro\tLo{\tY-0.10}\pgfmathsetmacro\tHi{\tY+0.10}%
    \pgfmathsetmacro\tWlo{\tE-\tSd}\pgfmathsetmacro\tWhi{\tE+\tSd}%
    \edef\temp{%
        \noexpand\fill[cCyc] (axis cs:0,\tLo) rectangle (axis cs:\tE,\tHi);
        \noexpand\whisker{\tWlo}{\tWhi}{\tY}
        \noexpand\node[cyclevalue] at ({rel axis cs:1.03,0} |- {axis cs:0,\tY})
            {= \noexpand\pgfmathprintnumber[fixed, fixed zerofill, precision=1, assume math mode=true]{\tE} \noexpand\textpm{} \noexpand\pgfmathprintnumber[fixed, fixed zerofill, precision=1, assume math mode=true]{\tSd}\noexpand\,ms};}%
    \temp}
\providecommand{\whisker}[3]{%
    \draw[black!70, line width=0.4pt] (axis cs:#1,#3) -- (axis cs:#2,#3)
        (axis cs:#1,#3-0.14) -- (axis cs:#1,#3+0.14) (axis cs:#2,#3-0.14) -- (axis cs:#2,#3+0.14);}
\providecommand{\tintswatch}[1]{\tikz[baseline=-0.6ex]{%
    \foreach \c [count=\i from 0] in {cDsp, cRec, cCat} {\fill[\c!#1] (\i*0.1cm,-0.07) rectangle ++(0.1cm,0.14);}}}

\begin{axis}[
    name=trace,
    scale only axis,
    width=4.20cm,
    height=2.50cm,
    xmin=0, xmax=135,
    ymin=-0.6, ymax=7.6,
    ytick={0, 1, 2, 3, 4, 5, 6, 7},
    yticklabels={\texttt{catalog}, \texttt{recommender}, \texttt{bci-processor}, \textbf{SASRec lifecycle},
                 \texttt{catalog}, \texttt{recommender}, \texttt{bci-processor}, \textbf{ACE lifecycle}},
    yticklabel style={font=\fontsize{6.0}{7.5}\selectfont},
    ytick pos=left,
    xtick pos=left,
    xtick={0, 25, 50, 75, 100, 125},
    xticklabel style={font=\fontsize{6.5}{7.5}\selectfont, /pgf/number format/assume math mode=true},
    xlabel={Time from the browser's request (ms)},
    xlabel style={font=\fontsize{6.5}{7.5}\selectfont, yshift=3pt},
    axis line style={gray!50, line width=0.4pt},
    tick style={gray!50, line width=0.4pt},
    xmajorgrids=true,
    grid style={color=cGrid, line width=0.4pt},
    tick align=outside,
    clip=false,
]
    \draw[dotted, gray!50, line width=0.4pt] (axis cs:0, 3.5) -- (axis cs:135, 3.5);
    \draw[densely dashed, black!65, line width=0.5pt] (axis cs:125, -0.6) -- (axis cs:125, 7.6);
    \node[rotate=90, anchor=south, inner sep=1.5pt, font=\fontsize{5.5}{6.5}\selectfont, black!75]
        at (axis cs:125, 3.5) {next EEG frame (@8\,Hz)};
    \node[anchor=south west, inner sep=0pt, font=\fontsize{5.5}{6.5}\selectfont, gray!80!black]
        at (rel axis cs:1.03,1) {send + proc + recv (ms)};

    \tracecycle{0}            
    \tracerow{0}{cDsp}        
    \tracerow{1}{cRec}        
    \tracerow{2}{cCat}        
    \tracecycle{1}            
    \tracerow{3}{cDsp}
    \tracerow{4}{cRec}
    \tracerow{5}{cCat}
\end{axis}

\node[anchor=north, inner sep=0pt, font=\fontsize{6}{7}\selectfont]
    at ($(current bounding box.south)+(0, -0.12cm)$)
    {\tintswatch{30}~Communication overhead (send / recv) \enspace
     \tintswatch{100}~microservice processing (proc) \enspace
     \tikz[baseline=-0.6ex]{\draw[black!70, line width=0.4pt] (0,0) -- (0.30cm,0) (0,-0.04cm) -- (0,0.04cm)
        (0.30cm,-0.04cm) -- (0.30cm,0.04cm);}~\textpm 1 sd};

\end{tikzpicture}
\caption{End-to-end latency breakdown and execution trace of a live recommendation cycle for ACE and SASRec.
The \textit{recommender} microservice dominates overall latency, while total cycle time remains strictly below the 125 ms deadline of the 8 Hz EEG stream, preventing queuing delays.}
\label{fig:performance}
\end{figure}
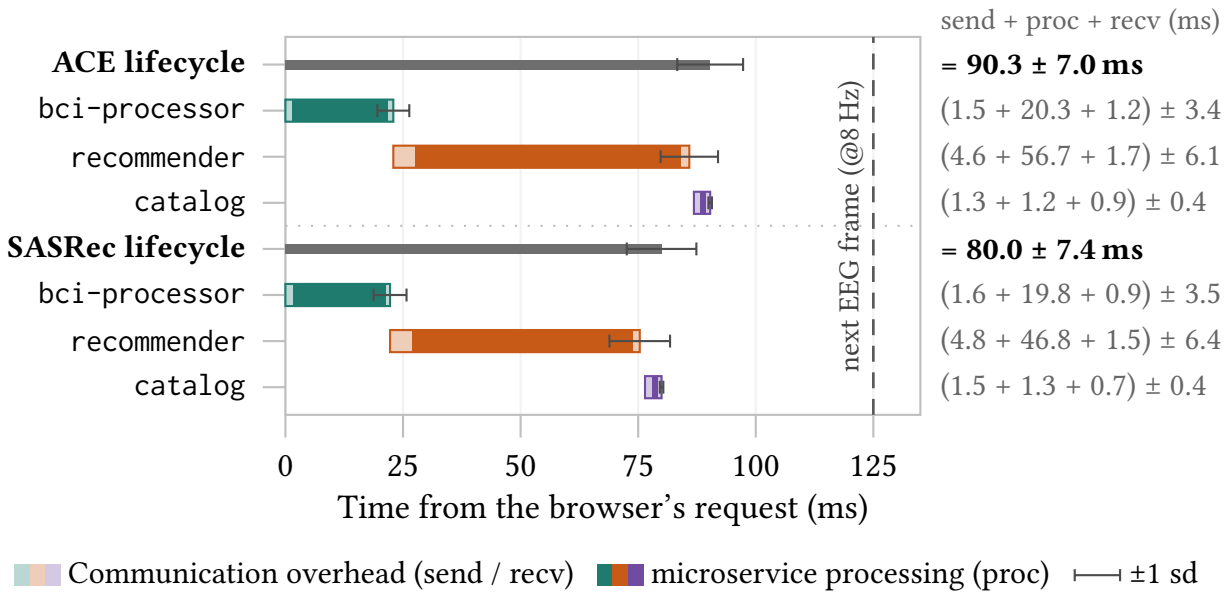

\newif\ifjourneycompact
\journeycompactfalse

\newlength{\jgap}\setlength{\jgap}{1.2mm}  
\newlength{\jH}                             
\newcounter{shot}

\newcommand{\shot}[3]{%
  \stepcounter{shot}%
  \begin{tikzpicture}[baseline=(img.south)]
    \node[inner sep=0pt] (img) {\includegraphics[height=#1, trim=#2, clip]{images/journey/#3}};
    \node[circle, draw=black!35, fill=white, inner sep=0pt, minimum size=2.7mm,
          font=\sffamily\bfseries\fontsize{6}{6}\selectfont, anchor=north east]
      at ([xshift=-0.8mm, yshift=-0.8mm]img.north east) {\alph{shot}};
  \end{tikzpicture}}

\newcommand{\jcol}[2]{\vbox{\offinterlineskip\hbox{#1}\vskip\jgap\hbox{#2}}}


\begin{figure*}[t]
  \centering
  \ifjourneycompact
    \setlength{\jH}{3.17cm}%
    \resizebox{\textwidth}{!}{%
      \jcol{\shot{\dimexpr(\jH-\jgap)*1284/1920\relax}{44 272 44 140}{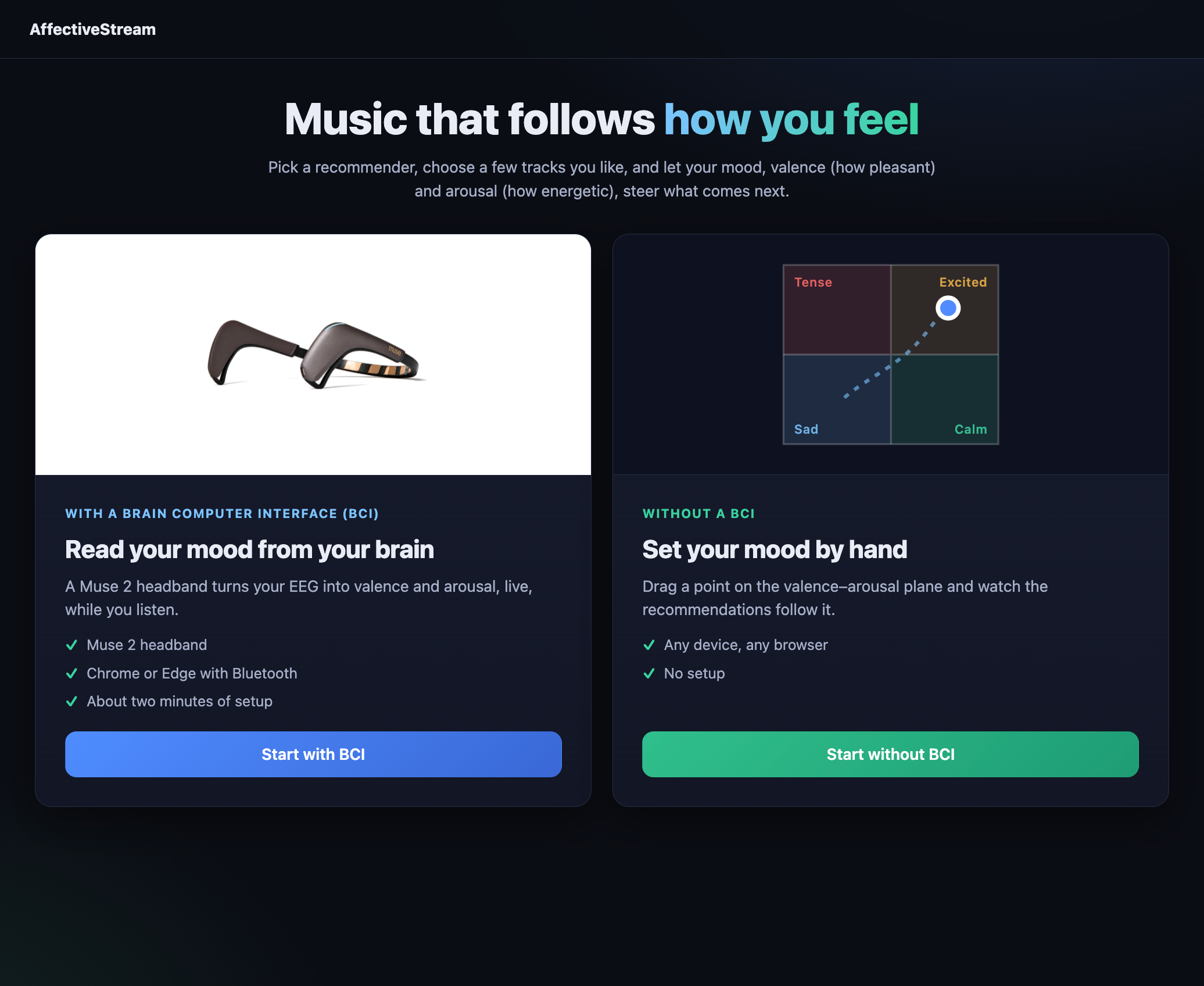}}%
           {\shot{\dimexpr(\jH-\jgap)*636/1920\relax}{44 854 44 206}{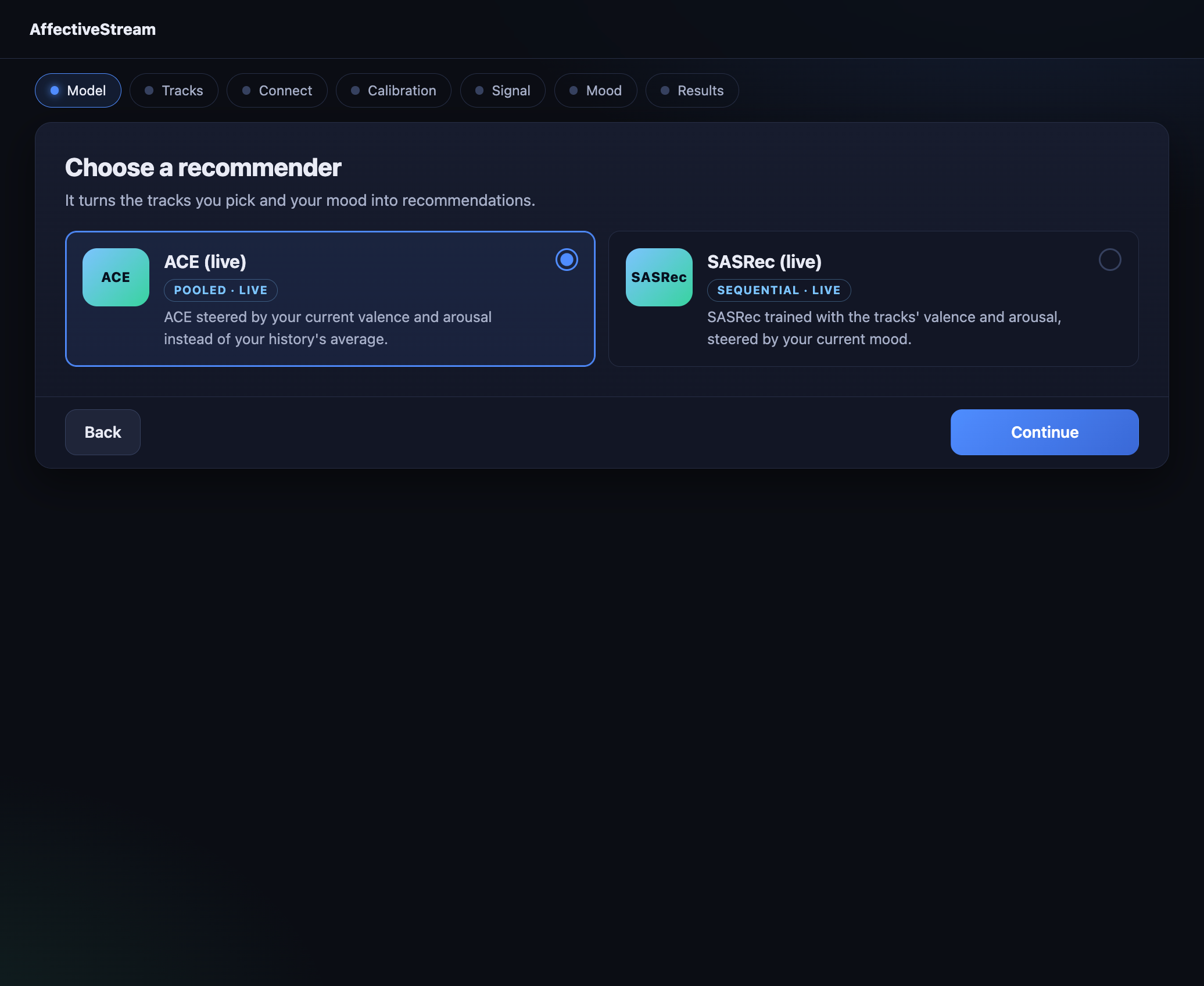}}%
      \hspace{\jgap}%
      \shot{\jH}{44 172 44 196}{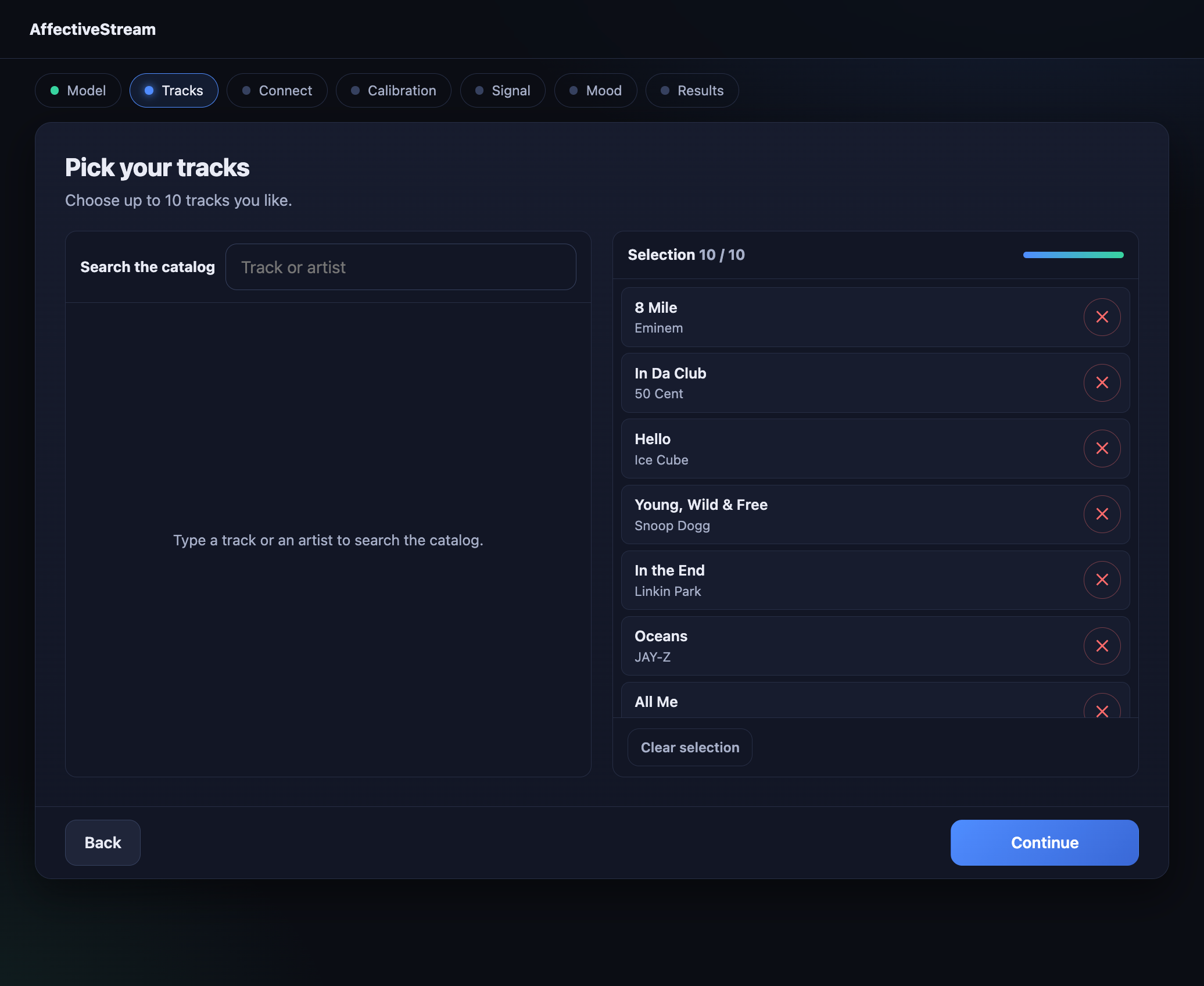}%
      \hspace{\jgap}%
      \shot{\jH}{44 334 44 196}{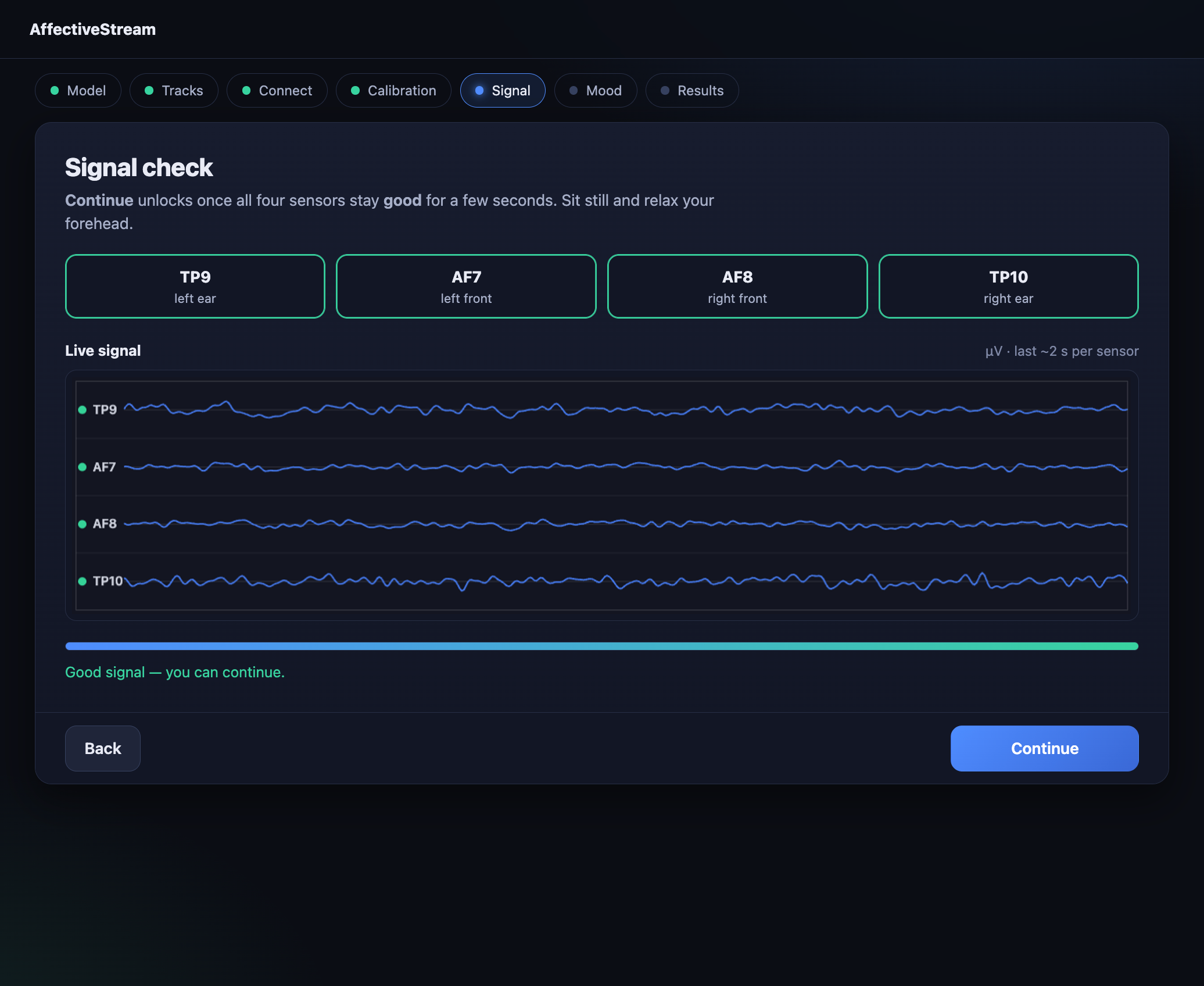}%
      \hspace{\jgap}%
      \shot{\jH}{44 0 44 196}{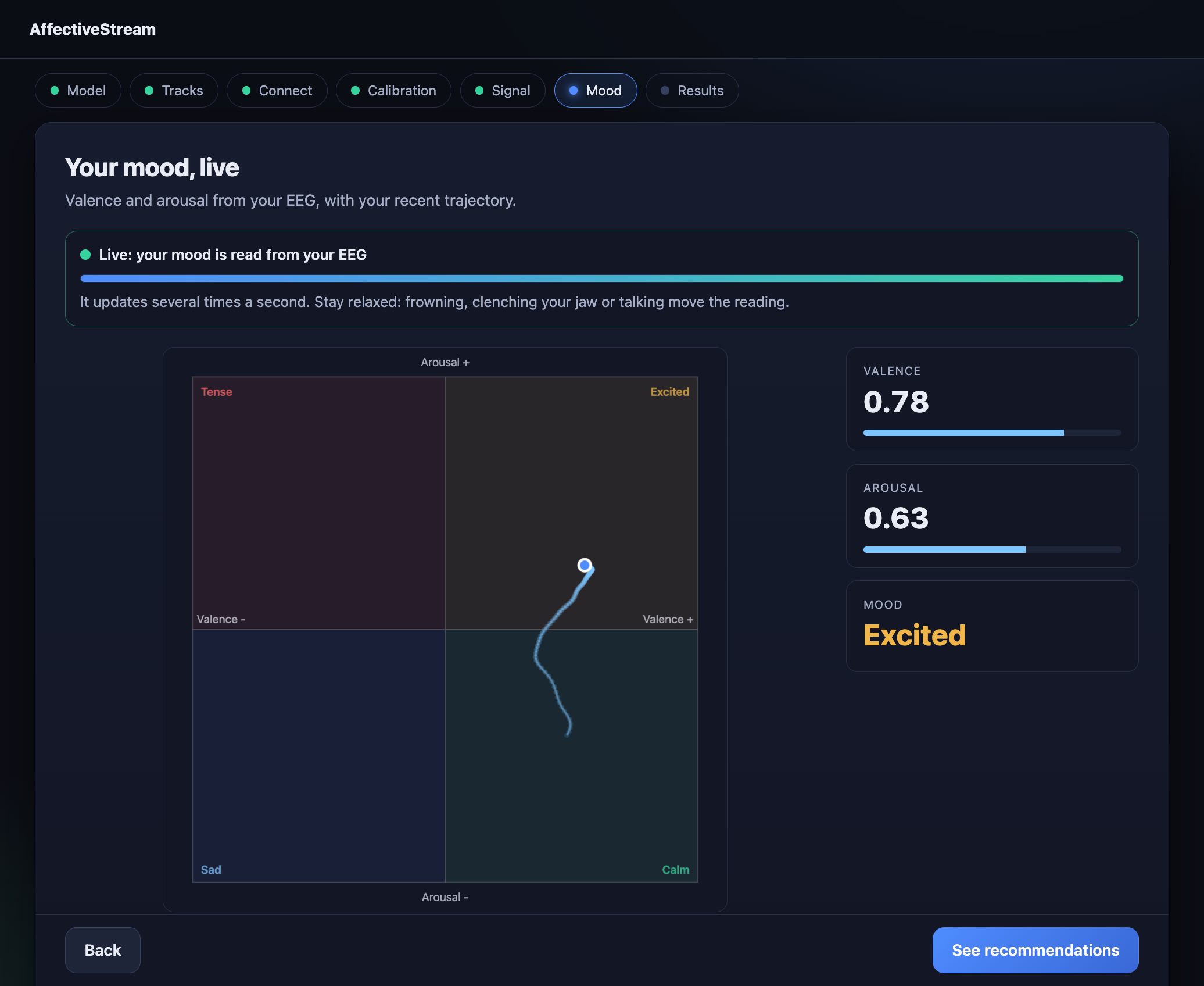}}
    \caption{The attendee's journey to the recommendation page: (a)~input modality, (b)~model, (c)~seed tracks, (d)~signal check, which in the BCI flow follows headband pairing and electrode placement, and (e)~live \va on Russell's plane. The manual flow skips (d) and replaces the live point with a draggable cursor.}
    \Description{Five screenshots of the web application: the landing page with the choice between the BCI and the manual flow, the model selection page, the seed-track selection page, the signal check page with per-channel EEG traces, and the mood page showing a moving point on the valence-arousal plane.}
  \else
    \setlength{\jH}{4.16cm}%
    \resizebox{\textwidth}{!}{%
      \jcol{\shot{\dimexpr(\jH-\jgap)*1284/1920\relax}{44 272 44 140}{landing}}%
           {\shot{\dimexpr(\jH-\jgap)*636/1920\relax}{44 854 44 206}{model}}%
      \hspace{\jgap}%
      \jcol{\shot{\dimexpr(\jH-\jgap)*1328/2142\relax}{44 172 44 196}{tracks}}%
           {\shot{\dimexpr(\jH-\jgap)*814/2142\relax}{44 686 44 196}{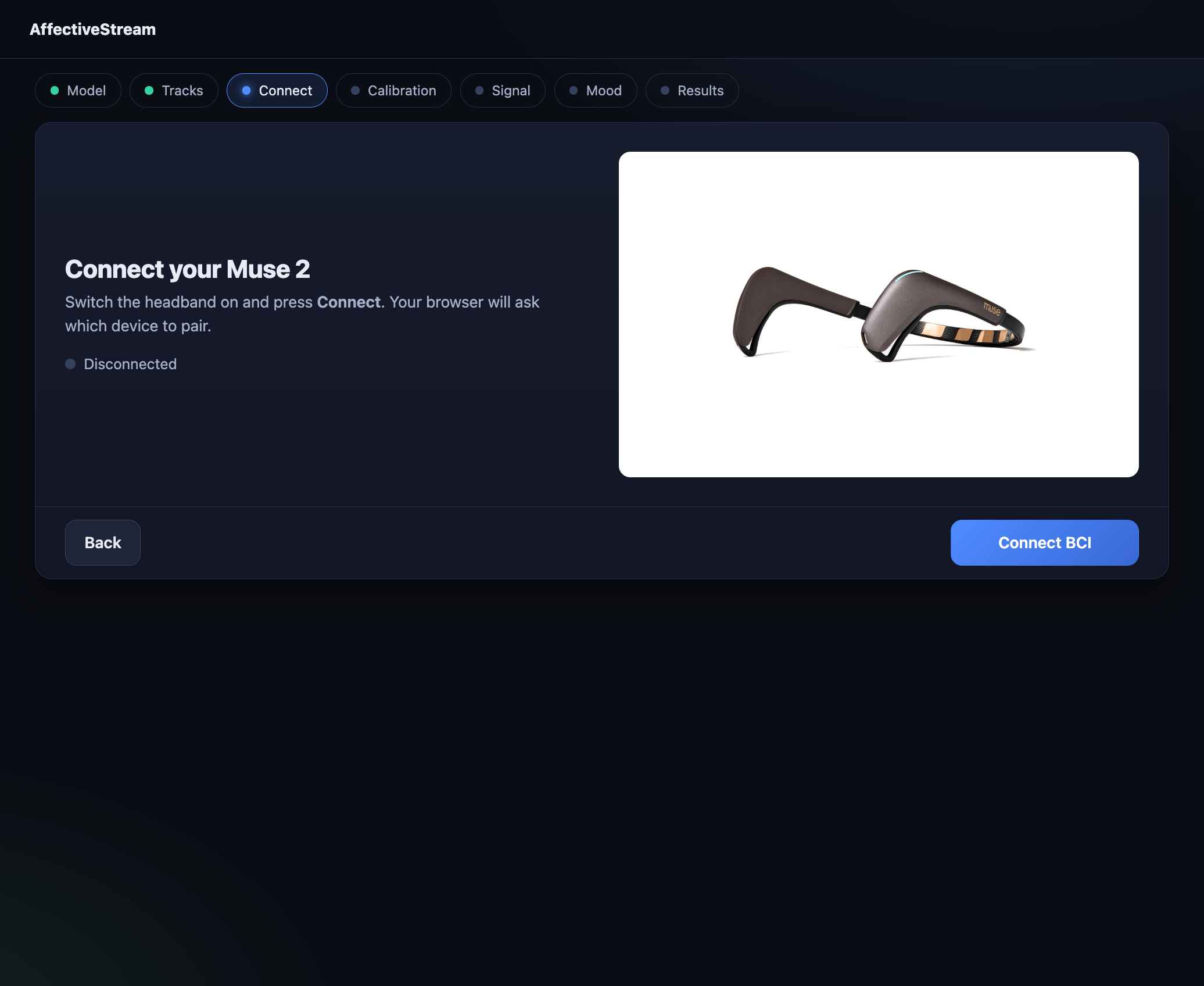}}%
      \hspace{\jgap}%
      \jcol{\shot{\dimexpr(\jH-\jgap)*814/1980\relax}{44 686 44 196}{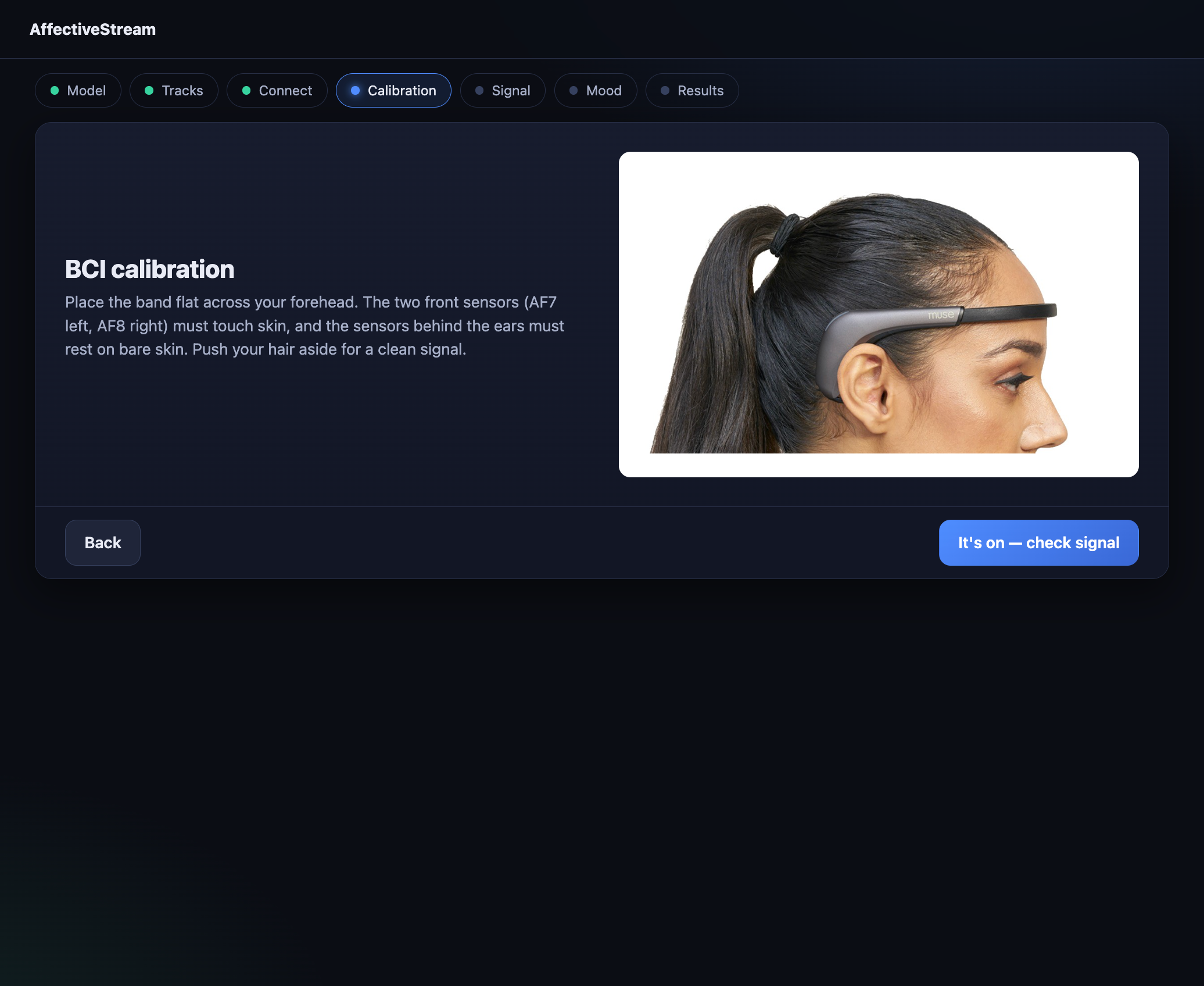}}%
           {\shot{\dimexpr(\jH-\jgap)*1166/1980\relax}{44 334 44 196}{signal}}%
      \hspace{\jgap}%
      \shot{\jH}{44 0 44 196}{mood}}
    \caption{BCI Walkthrough: (a)~input selection, (b)~recommender model, (c)~favorite tracks selection, (d)~device pairing, (e)~sensor placement, (f)~signal quality check, and (g)~live \va\ tracking. The manual mode skips (d)-(f) via a draggable 2D cursor.}
    \Description{Seven screenshots of the web application: the landing page with the choice between the BCI and the manual flow, the model selection page, the seed-track selection page, the headband pairing page, the electrode placement page, the signal check page with per-channel EEG traces, and the mood page showing a moving point on the valence-arousal plane.}
  \fi
  \label{fig:journey}
\end{figure*}

\begin{figure}[t]
  \centering
  \includegraphics[width=\columnwidth, trim=44 0 44 196, clip]{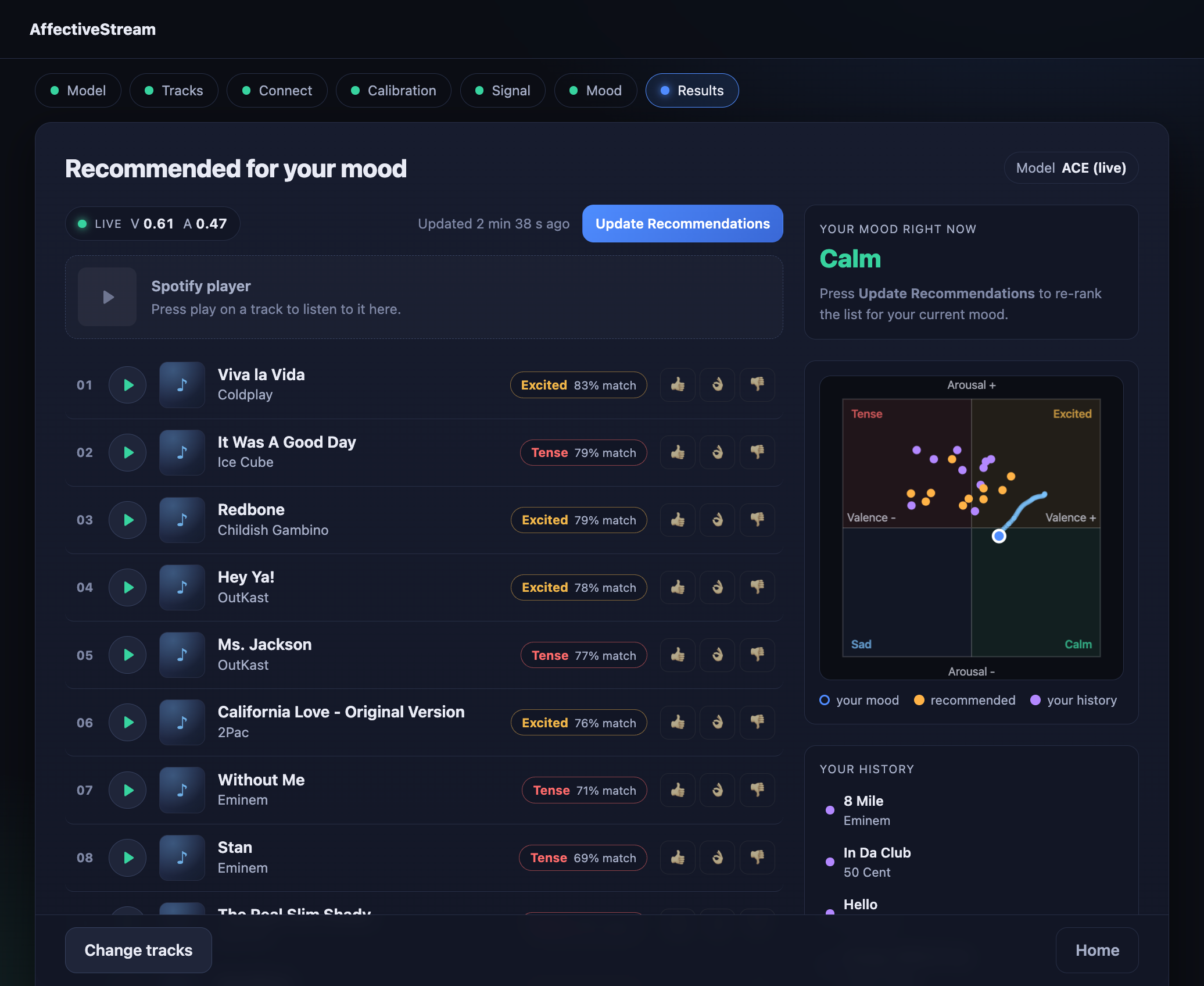}
  \caption{Recommendation page: top-10 V/A-conditioned tracks with audio playback and ratings. The plane displays live \va\ proxy coordinates, with on-demand re-ranking.}
  \label{fig:recommendations}
\end{figure}

To isolate platform performance from network delays, we benchmarked the system locally on a MacBook Pro M4 via WarpRec~\cite{DBLP:journals/corr/abs-2602-17442} (Figure~\ref{fig:performance}).
The system serves the Lastfm-VADS catalog, which provides inferred item-level V/A profiles~\cite{DBLP:conf/sigir/AghilarRABTN26}. Although the figure illustrates that candidate scoring in the \textit{recommender} microservice dominates latency, both \textsc{ACE} and \textsc{SASRec} (extended with the \textit{affective adapter}) complete well below the $125~\text{ms}$ deadline of the $8~\text{Hz}$ \ac{EEG} stream, confirming that the architecture operates without buffer accumulation or queuing delays. Finally, serving models via the affective adapter preserves recommendation quality, yielding no statistically significant difference (Wilcoxon signed-rank test, $p{>}0.05$) in $\text{nDCG@10}$ between live and offline counterparts: $0.0231$ (live) vs. $0.0236$ (offline) for \textsc{ACE}~\cite{DBLP:conf/sigir/AghilarRABTN26}, and $0.0164$ (live) vs. $0.0148$ (offline) for \textsc{SASRec}~\cite{DBLP:conf/icdm/KangM18}.

\section{Demonstration and Scenarios}
\label{sec:demonstration}

The demonstration invites attendees to experience the system through an interactive web application (Figures~\ref{fig:journey} and~\ref{fig:recommendations}). We show the system across two distinct operational scenarios, highlighting both \ac{BCI} interaction and a hardware-free alternative.

\subsection{Scenario 1: Interactive BCI Walkthrough}
\label{sec:scenario_bci}

In this primary scenario, attendees experience the complete closed-loop interaction driven by consumer \ac{EEG}. First, the visitor configures their listening profile by choosing a recommendation model between \textit{ACE (live)} and \textit{SASRec (live)} and selecting up to ten favorite tracks from the catalog to initialize preferences (Figure~\ref{fig:journey}a-c).
Next, the web application connects to a Muse~2 headband over the Web Bluetooth API, displays real-time electrode contact check, and verifies signal stability through the quality gate (Figure~\ref{fig:journey}d-f). Once the signal stabilizes, the system continuously  maps the incoming \ac{EEG} stream into normalized V/A proxy coordinates, projecting them onto Russell’s 2D circumplex plane
at 8\,Hz (Figure~\ref{fig:journey}g).
Finally, the system captures the current \va coordinates and generates a refreshed top-10 recommendation list (Figure~\ref{fig:recommendations}). Attendees can listen to V/A-conditioned tracks through the embedded player and record three-tier ratings (liked, indifferent, disliked). 
Explicit session resets clear all in-memory buffers, WebSocket connections, and user data.

\subsection{Scenario 2: Hardware-Free Exploration}
\label{sec:scenario_manual}

This secondary scenario demonstrates an alternative interaction mode that operates entirely without specialized \ac{EEG} hardware. The attendee selects the manual flow, bypassing the sensor pairing and calibration screens to navigate Russell's 2D plane directly with an interactive cursor. Direct manipulation of this 2D cursor across different emotional quadrants (such as calm relaxation or high-arousal focus) enables visitors to inspect how the ranking models dynamically re-score the catalog in real time. This capability is guaranteed by the system's input swappability feature: the architecture accommodates interchangeable input modalities without downstream modifications, provided that the exchanged data format for valence and arousal coordinates remains identical.

\section{Conclusion}
\label{sec:conclusion}

Real-time affective music recommendation has long been constrained by two practical barriers: wearable devices tied to third-party drivers, and recommendation pipelines that depend on static user histories. To overcome both limitations, our web platform pairs consumer \ac{EEG} hardware directly with the browser via Web Bluetooth. By mapping raw signals onto Russell's valence-arousal plane, the architecture decouples physiological processing from candidate scoring. This separation enables an affective adapter to serve anonymous visitors on the fly across both graph neural networks and sequential transformer backbones. Under a shared live-affect scoring mechanism, the module infers musical preferences from a few favorite tracks and balances them against the live EEG-derived control signal, bypassing static user embeddings.


\section{Ethical Statement}
\label{sec:ethics}

All \ac{EEG} signals, affective coordinates, and preferences are processed ephemerally in memory and deleted after each session.
The platform uses affective telemetry as a control signal to steer recommendations, rather than for validated emotion recognition or a clinical assessment of emotional health.





\bibliographystyle{ACM-Reference-Format-initials}
\bibliography{main}

\end{document}